**Nonlinear performance of asymmetric coupler based on dual-core photonic crystal fiber: towards sub-nanojoule solitonic ultrafast all-optical switching**

**L. Curilla[a,*], I. Astrauskas[b], A. Pugzlys[b], P. Stajanca[c], D. Pysz[d], F. Uherek[e], A. Baltuska[b], I. Bugar[d]**

[a] Department of Experimental Physics, Faculty of Mathematics, Physics and Informatics, Comenius University, Mlynska dolina, 842 48 Bratislava, Slovak Republic
[b] Photonics Institute, Vienna University of Technology, Gusshausstrasse 27-387, 1040 Vienna, Austria
[c] Federal Institute for Materials Research and Testing, Unter den Eichen 87, 12205 Berlin, Germany
[d] Institute of Electronic Materials Technology, Wolczynska 133, 01-919 Warsaw, Poland
[e] International Laser Centre, Ilkovicova 3, 841 04 Bratislava, Slovak Republic

**Abstract**
We demonstrate ultrafast soliton-based nonlinear balancing of dual-core asymmetry in highly nonlinear photonic crystal fiber at sub-nanojoule pulse energy level. The effect of fiber asymmetry was studied experimentally by selective excitation and monitoring of individual fiber cores at different wavelengths between 1500 nm and 1800 nm. Higher energy transfer rate to non-excited core was observed in the case of fast core excitation due to nonlinear asymmetry balancing of temporal solitons, which was confirmed by the dedicated numerical simulations based on the coupled generalized nonlinear Schrödinger equations. Moreover, the simulation results correspond qualitatively with the experimentally acquired dependences of the output dual-core extinction ratio on excitation energy and wavelength. In the case of 1800 nm fast core excitation, narrow band spectral intensity switching between the output channels was registered with contrast of 23 dB. The switching was achieved by the change of the excitation pulse energy in sub-nanojoule region. The performed detailed analysis of the nonlinear balancing of dual-core asymmetry in solitonic propagation regime opens new perspectives for the development of ultrafast nonlinear all-optical switching devices.

**1. Introduction**

Photonic crystal fiber (PCF)[1] research after its great advancement in the last two decades still bears inspiring innovation potential in various disciplines of applied optics such as signal processing and telecommunications. Especially, multi-core PCFs are exciting media for temporal-spatial-spectral optical field transformations due to the possibility of tailoring of their dispersion, coupling and nonlinear optical properties by proper microstructure engineering [1]. Thanks to the coupling between optical fields in its individual cores, dual-core PCFs (DC PCFs) can be used as couplers (for example as multiplexers [2]) or all-optical switches [2]. According to the supermode theory, the key parameter characterizing the coupling phenomenon is the coupling length $L_C$, representing the shortest distance at which the signal is transferred from one core to another in a periodical manner. The coupling length is defined by the formula

---

[1] *List of abbreviations:*
DC PCF – Dual-Core Photonic Crystal Fiber; IR – InfraRed; NLDC – NonLinear Directional Coupler; SEM – Scanning Electron Microscope; CNLSE – Coupled NonLinear Schrödinger Equation; NLSE – NonLinear Schrödinger Equation; OPA – Optical Parametric Amplifier; ER – Extinction Ratio

$$L_C = \frac{\pi}{|\beta_S - \beta_A|}, \quad (1)$$

where $\beta_S$ and $\beta_A$ are propagation constants of the symmetric and anti-symmetric supermode, respectively. Supermodes are the fundamental eigenmodes of a dual-core fiber determined by the fiber material and microstructure [3]. The second important parameter of a dual-core fiber coupler is the inter-core coupling represented by an effective coupling coefficient $\kappa_e$, depending on the measure of optical asymmetry between the two fiber cores. The higher the asymmetry, the less efficient is the energy transfer between the cores. Effective coupling coefficient $\kappa_e$ is defined as

$$\kappa_e = \sqrt{\kappa_{12} \cdot \kappa_{21} + \delta^2}, \quad (2)$$

where $\kappa_{ij}$ describes coupling from $i$-th core to $j$-th core and $\delta$ is propagation constant mismatch between the cores representing the asymmetry. Propagation constant mismatch $\delta$ is defined as $\delta = (\bar{\beta}_2 - \bar{\beta}_1)/2$, where $\bar{\beta}_{1,2}$ represent the propagation constants in the individual cores [3].

In practice, it is very difficult to produce ideally symmetric DC PCF [4] and the cores of a dual-core fiber exhibit slightly different effective refractive indices $n_{eff}$ causing mismatch of propagation constants. Due to the asymmetry the two cores are distinguishable on the base of the propagation constant determining slow ($\bar{\beta}_1$) and fast ($\bar{\beta}_2 < \bar{\beta}_1$) core. The asymmetry prevents effective energy transfer between them, when

$$\delta \geq \frac{\sqrt{\pi^2 - 4L_C^2 \kappa_{12}\kappa_{21}}}{2L_C} \quad (3).$$

Let us consider positive Kerr nonlinearity typical for the majority of glass materials. In this case, launching pulse with suitable peak-power into the fast core, Kerr-induced phase shift decreases the mismatch between the cores and thus causes balancing of the dual-core asymmetry. The situation reverses when the slow core is excited at the high-power levels: the asymmetry between the cores increases due to the Kerr effect and the cores may become practically uncoupled.

Due to the small inter-core distances enabled by the PCF technology, $L_C$ in the millimeter-range is easily accessible allowing the application of shorter fiber couplers in comparison to the classical dual-core fibers. Coupling properties of dual-core fibers exhibit birefringent character. Thanks to that, control of the intensity ratio between the two output ports (dual-core extinction ratio) by alternating the input polarization direction is possible [5, 6]. In nonlinear regime, DC PCFs can be used as nonlinear directional couplers (NLDC) and represent an interesting alternative for all-optical signal processing applications. In the case of single core excitation of ideally symmetric DC PCF with length equal to $L_C$, low power pulses undergo periodic coupling and exit the fiber from the non-excited core. In contrast, high power pulses with proper peak-intensity level exit predominantly from the excited core because the nonlinearly-induced asymmetry decreases the inter-core coupling efficiency. However, the nonlinear propagation is often associated with pulse deformations, which may deteriorate the nonlinear switching process. Therefore, soliton propagation regime in anomalous dispersion region was suggested for establishing more effective NLDC performance. According to the theory, temporal solitons with proper characteristics remain stable during oscillations between the cores of a dual-core fiber [7], eliminating not only the dispersion, but also the dual-core propagation caused perturbations. Employing PCF structures, numerous theoretical works analyzed the soliton-based NLDC possibilities, e.g. by designing asymmetric couplers [8], utilizing highly nonlinear glass [9] or suggesting Tbit/s rate all-optical logic components [10, 11]. On the other hand, few experimental attempts motivated by the vision of fiber based solitonic NLDC were unsuccessful, mainly due to hardly controllable process of soliton fission, i.e. soliton temporal break-up into its fundamental components, which is inherently present during the nonlinear propagation in anomalous dispersion region [2, 12].

Soliton fission is a generally adopted propagation scenario in the case of highly nonlinear PCFs in the anomalous dispersion region [13, 14]. According to this concept, high order solitonic pulse breaks up into its fundamental components influenced by the high order perturbation effects. Moreover, subsequently the fundamental components undergo further spectral transformations due to Kerr and Raman nonlinearities typically resulting in broad and structured output spectra, whose complexity depends on the order of the initial soliton. In the case of propagation of low-order soliton in slightly birefringent PCF, nonlinear coupling between the two orthogonal polarization components has been studied both theoretically and experimentally [15]. Both approaches confirmed the possibility of energy transfer from the fast to slow polarization component without observation of the opposite effect. This concept is straightforwardly adaptable for the coupling between fast and slow core of an asymmetric dual-core coupler. However, dual-core technology promises to demonstrate similar effect at significantly shorter fiber lengths in comparison to [15], where 2 m of the birefringent PCF was necessary.

Experimental NLDC study using DC PCF made of standard silica was for the first time reported in 2006 [2]. 9 mm long fiber was excited by 120 fs pulses with central wavelength of 1550 nm, localized in the anomalous dispersion region of the fiber. Obtained results revealed only unidirectional switching increasing the dual-core extinction ratio from -11.0 dB to -1.8 dB level applying tens of nanojoule pulse energies. Later on, a nonlinear narrow band true switching has been demonstrated, both in visible [16] and NIR spectral regions [17] at non-excitation wavelengths. The utilized special DC PCF possessed a square microstructure design and was made of soft silicate glass with linear and nonlinear refraction indices similar to standard silica. The fiber had a slight dual-core asymmetry, which altered the switching character depending on the choice of the excited core but its impact on the switching mechanism was negligible [18]. The best switching performance has been achieved with 1650 nm 100 fs excitation pulses using 14 mm long piece of the fiber. The switching wavelengths were shifted about 100 nm from the excitation one [19]. The best switching contrast has been at the level of 15 dB (7 dB vs. -8 dB), which is promising also from the application point of view. On the other hand, demonstrated switching required relatively high pulse energies (tens of nanojoules). Therefore, fiber material with higher nonlinear refractive index would be suitable for achieving switching at lower energies.

Following up on this direction, new generation of DC PCFs was manufactured from similar silicate soft glass however with about 20-times stronger Kerr nonlinearity. At the same time, the fiber possess higher degree of dual-core asymmetry as well. Utilizing these novel specialty fibers, we investigate nonlinear enhancement of the dual-core coupling in femtosecond solitonic propagation regime. Basic optical properties of the novel DC PCFs are presented, followed by the first experimental results of nonlinearly-induced spatial-spectral pulse transformations in these fibers. Excitation wavelength is varied in the 1500 – 1800 nm spectral region covering the flattened dispersion region (Fig. 1b) in order to study the wavelength effect on the nonlinear dual-core propagation. Experimental findings are supported by the results of dedicated numerical simulations revealing the spectral peculiarities of the studied inter-core energy transfer process.

## 2. Characterization of the fibers and the methods of their investigation
### 2.1 Employed fiber characteristics

A new generation DC PCF with hexagonal microstructure and enhanced nonlinear performance potential (Fig. 1a) was self-designed [20]. In–house synthesized highly nonlinear glass (PBG08) with well determined linear and nonlinear properties was used for fabrication of the fiber samples in correspondence to the theoretical optimization process. PBG08 is a lead-bismuth-gallium-oxide glass with composition $SiO_2 - Bi_2O_3 - PbO - Ga_2O_3 - CdO$ and with improved transmission in the NIR spectral region. Nonlinear refractive index of this soft glass is $n_{NL} = 4.3 \times 10^{-19} m^2/W$, which is almost 20-times higher than that of fused silica and one of the highest values among other oxide glasses [21]. In addition, PBG08 has more than three-times lower relative Raman contribution coefficient $f_R$ as compared to silica glass, which is beneficial for the stability of dissipative solitons [22].

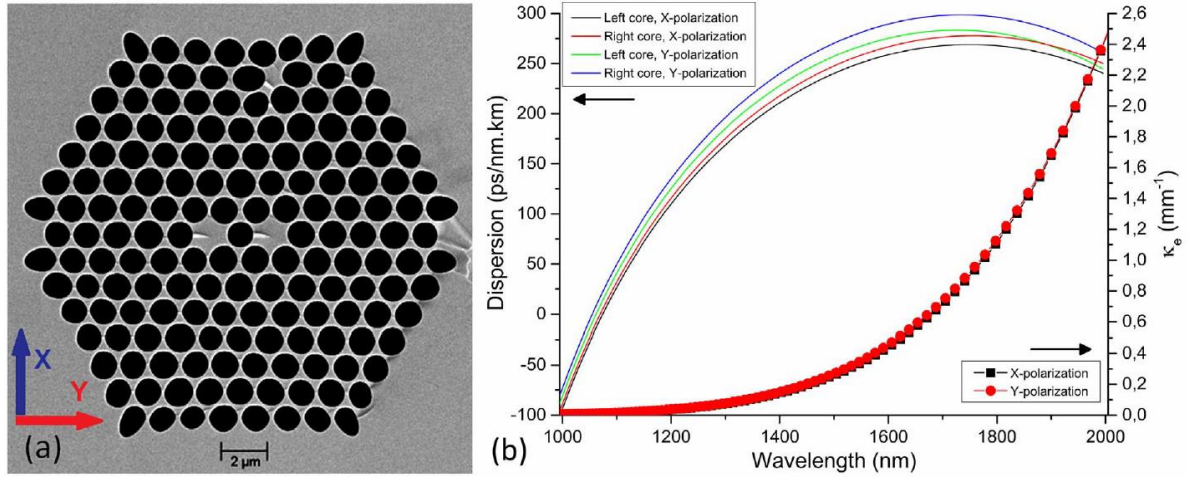

Fig. 1. (a) Scanning electron microscope (SEM) image of the cross-section of the investigated asymmetric DC PCF microstructure (NL37D sample) with indication of X and Y polarization directions. (b) Numerically calculated dispersion curves of the NL37D single-core modes and the corresponding effective coupling coefficient $\kappa_e$ spectral characteristics for the X- and Y-polarization.

Two fiber samples with slightly different structure were prepared as outcomes of two separate fiber drawing procedures. The samples are designated as NL37C and NL37D and their geometrical parameters are presented in Table 1. The core size values in the table are referred for the both cores, because they are indistinguishable taking into account the uncertainty of their determination.

| Sample | Diameter | Microstructure dimension | Lattice pitch | Air hole diameter | Core size | |
|---|---|---|---|---|---|---|
| | | | | | Y-axis | X-axis |
| **NL37C** | 119 μm | 20.22 μm | 1.4 μm | 1.2 μm | 1.55 μm | 1 μm |
| **NL37D** | 110 μm | 18.7 μm | 1.2 μm | 1 μm | 1.42 μm | 0.9 μm |

Table 1. Geometrical properties of the two studied fiber samples.

After the fabrication, simulation of basic linear characteristics was performed in the case of both samples based on the scanning electron microscope images and dispersion of PBG08. For that purpose, a commercial mode solving software was used, that enables to compute the dispersion characteristics of the fiber supermodes. Additionally, individual core characteristics were calculated, putting an artificial air hole at the place of the opposite core preserving dimensions and periodicity of the surrounding air holes. Since DC PCFs exhibit also birefringent behavior, it is important to consider the polarization aspect. Therefore, X- and Y-polarization denotes polarization state with electric field perpendicular and parallel to the connecting line of the two cores (Fig. 1a). Non-negligible asymmetry of the both samples was revealed during the simulation work. Therefore, it is necessary to distinguish between the cores in terms of fast and slow one. The slow core is situated on the left and the fast core on the right side of the image in Fig. 1a. Linear propagation parameters of the individual cores as determined by mode solver at wavelength 1600 nm are summarized in Table 2.

| Sample | $n_{eff}$ @ 1600 nm | | | | $\bar{\beta} \times 10^3$ (1/mm) @ 1600 nm | | | |
|---|---|---|---|---|---|---|---|---|
| | X-polarization | | Y-polarization | | X-polarization | | Y-polarization | |
| | slow core | fast core | slow core | fast core | slow core | fast core | slow core | fast core |
| **NL37C** | 1.7488 | 1.7462 | 1.7453 | 1.7425 | 6.8640 | 6.8538 | 6.8503 | 6.8393 |
| **NL37D** | 1.7348 | 1.7337 | 1.7316 | 1.7299 | 6.8091 | 6.8048 | 6.7965 | 6.7899 |

Table 2. Basic optical properties of two studied fiber samples at the wavelength of 1600 nm.

In case of NL37D, comparison of the propagation constants of slow and fast core yields to dual-core asymmetry of $\delta_D^X = 2.15$ $mm^{-1}$ for X-polarization and $\delta_D^Y = 3.3$ $mm^{-1}$ for Y-polarization. Roughly two times larger asymmetry was determined in case of NL37C sample ($\delta_C^X = 5.1$ $mm^{-1}$, $\delta_C^Y = 5.5$ $mm^{-1}$). These values reflect also birefringent character of the asymmetry because it depends not only on the dimensions of the cores, but also on the surrounding structure. All parameters in Table 2 were calculated for wavelength 1600 nm, because important experimental results were obtained under femtosecond excitation at this central wavelength. However the spectral dependences of all linear parameters were acquired by using the mode solver, too.

These characteristics are very similar in the case of both samples, thus Fig. 1b shows fiber dispersion and effective coupling coefficient $\kappa_e$ dependence on wavelength only for the NL37D sample. Due to the asymmetry, the effective coupling between the cores of both fiber samples is very weak in linear regime resulting low values of $\kappa_e$ (less than 1 mm$^{-1}$ around 1600 nm). However, it significantly increases with increasing wavelength, which is clearly observable on Fig. 1b, thus the longer wavelengths experience weaker asymmetry. From the point of view of our topics it means that lower field intensity is required at longer wavelength for balancing the asymmetry by nonlinear interaction in comparison to the shorter ones. Thanks to the high nonlinearity of the fiber and to the ultrafast pulse excitation, the negative impact of asymmetry can be eliminated already at relatively low pulse energies (sub-nJ) and the balancing energy decreases monotonically with increasing wavelength.

*2.2 Theoretical concept and nonlinear simulations*
The optical soliton propagation and switching in NLDC is usually modeled by a pair of coupled nonlinear Schrödinger equations (CNLSE) [22–24]. We have modified in the past [25] the general form of these equations involving non-negligible effects, those are emerging in dual-core PCFs. More generalized form of the CNLSE was derived, taking into consideration the coupling coefficient dispersion, losses, stimulated Raman scattering and self-steepening effect, that induce group velocity mismatch between the pulse peak and wings leading to a steepened slope of its trailing part. Later on, the model was extended by inclusion of cross-phase modulation effect and spectral dependence of the self-steepening nonlinearity, which led to improvement of qualitative agreement with experimental results presented in work [19]. The model represents an appropriate tool for addressing the asymmetry effects, because the linear modal characteristics are calculated and implemented separately for both cores. The dispersion characteristics of the fundamental modes of the individual fiber cores were calculated according to the above introduced approach supposing two complementary single-core structures. Experimentally determined instantaneous Kerr and delayed Raman responses of the PBG08 glass [21] are included into the material nonlinear response function. Two coupled equations were solved numerically by utilizing the split-step Fourier method enabling efficient transition between the frequency and time domain, performed at every step of the iterative field evolution calculation. Linear step of the solving routine is performed in the frequency domain where the full dispersion of propagation constant and effective coupling coefficient is considered. Linear characteristics of the fiber necessary for solving the CNLSE were evaluated in the spectral range of 300–3000 nm, sufficiently covering the spectral region of our interest.

Additionally, fiber losses $\alpha_{dB}$ and the in-coupling efficiency $C$ were evaluated experimentally by cutback method at wavelength 1550 nm and resulted in values $\alpha_{dB} = 27.7 \pm 1.2\ dB/m$ and $C = 32 \pm 1\%$, respectively. These values are valid for both fiber cores and for both main polarization directions in the case of the both fiber samples at the level of the determination uncertainty. In the next step, the spectral dependence of the losses in range of 300–3000 nm was defined as a sum of numerically calculated waveguide and measured material absorption losses scaled to the experimentally obtained value at 1550 nm. Nonlinear simulations using the presented model were performed subsequently for both X- and Y- input polarizations. Propagation distance of 10 mm was used for all the simulations in order to cover the lengths of fiber samples used in the experiment. Excitation pulses were assumed to have hyperbolic secant pulse-shape and duration of 100 fs. Input pulse energies in range of 0.5 - 10 nJ with step of 1 nJ above the 1 nJ level in correspondence to the experimental conditions were applied during the numerical simulations. The energy values were reduced by the above introduced in-coupling efficiency factor $C = 32\%$, which represents the ratio of the energy with respect to the energy measured before the coupling optics. Simulation series were repeated at different excitation wavelengths in the range of 1500 nm – 1800 nm with 100 nm steps.

*2.3 Experimental setup*

Dual-core photonic crystal fiber samples were studied using the experimental setup depicted in Fig. 2.

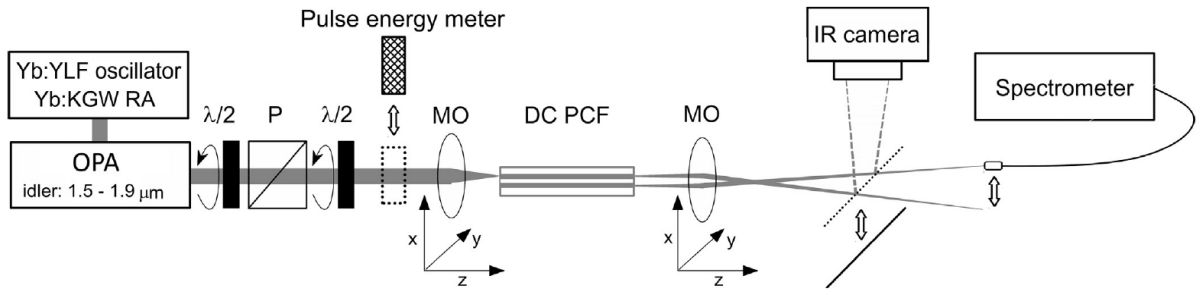

Fig. 2. Experimental setup used for study of spatial-spectral field transformations during ultrafast dual-core propagation in asymmetric DC PCFs. RA – regenerative amplifier, OPA – optical parametric amplifier, λ/2 – half-wave plate, P – polarizer; MO – micro-objective.

The idler output of a two-stage optical parametric amplifier (OPA) was used in the experiments as a source of wavelength-tunable near-infrared excitation pulses. OPA was driven by the second harmonics of Yb:KGW chirped pulse regenerative amplifier (RA) seeded by stretched output of Yb:YLF oscillator. The RA operates at 1030 nm central wavelength and produces 150 µJ, 250 fs pulses at 3 kHz repetition rate. The OPA generated idler pulses at energy level of 3 µJ with tuneability range 1500 - 1900 nm and with estimated duration of 100 fs, taking into consideration the calculated bandwidth-limited duration to be 70 fs.

Although the OPA was designed for operation in the spectral range between 1500 nm and 1900 nm, covering the flattened dispersion region of the studied DC PCFs (Fig.1b), 1800 nm excitation wavelength was used as the longest in the experiments due to the low sensitivity of the infrared (IR) camera above this wavelength. The advantage of excitation in the flat area of the fiber dispersion peak is moderate effect of third order dispersion. This approach eliminates unwanted broadband pulse deformations, such as dispersive wave generation which also causes further losses during the soliton fission process [26].

In order to control the energy of the input pulses generated by the OPA, the beam passed through an adjustable attenuator made of a half-wave plate ($\lambda/2$) and polarizer (P). A subsequent second half-wave plate was used to control the polarization direction and a micro-objective (40x) to couple the laser radiation into one of the fiber cores. In order to register spectra selectively from the individual cores, an achromatic micro-objective (50x) was used to image the output facet of the DC PCF on the input surface of the collection fiber of Ocean Optics Near-Quest spectrometer operating in the spectral range of 1100 – 2500 nm and having spectral resolution of ~3.3 nm. With a help of an auxiliary mirror, spatial field distribution of the DC PCF output facet was recorded by the IR camera after every change of the excitation conditions. Both micro-objectives were mounted on XYZ translation stages having sub-micrometer precision. The spatial-spectral intensity distribution at the output of the dual-core fiber cores was studied by varying the experimental conditions, similarly in the case of the both fiber samples. Two different fiber lengths (6 and 10 mm) were examined, following the strategy to approach sub-cm scale, but also to preserve uncomplicated sample handling. Both cores were selectively excited with pulse energies in the range of 0.5 nJ - 10 nJ measured before the input micro-objective. All the measurements were performed for both prominent orthogonal input polarization directions.

## 3. Results and discussion
### *3.1 Observation of nonlinear asymmetry balancing*

Firstly, we present experimentally observed nonlinear asymmetry balancing effect in DC PCF according to the IR camera records. Fig. 3 depicts images of spatial intensity distribution at the output facet of 10 mm long NL37D sample excited by Y-polarized 1600 nm pulses at different energy levels. They represent the best results from the point of view of dual-core extinction ratio change with increasing pulse energy according to the camera records among the all examined experimental series.

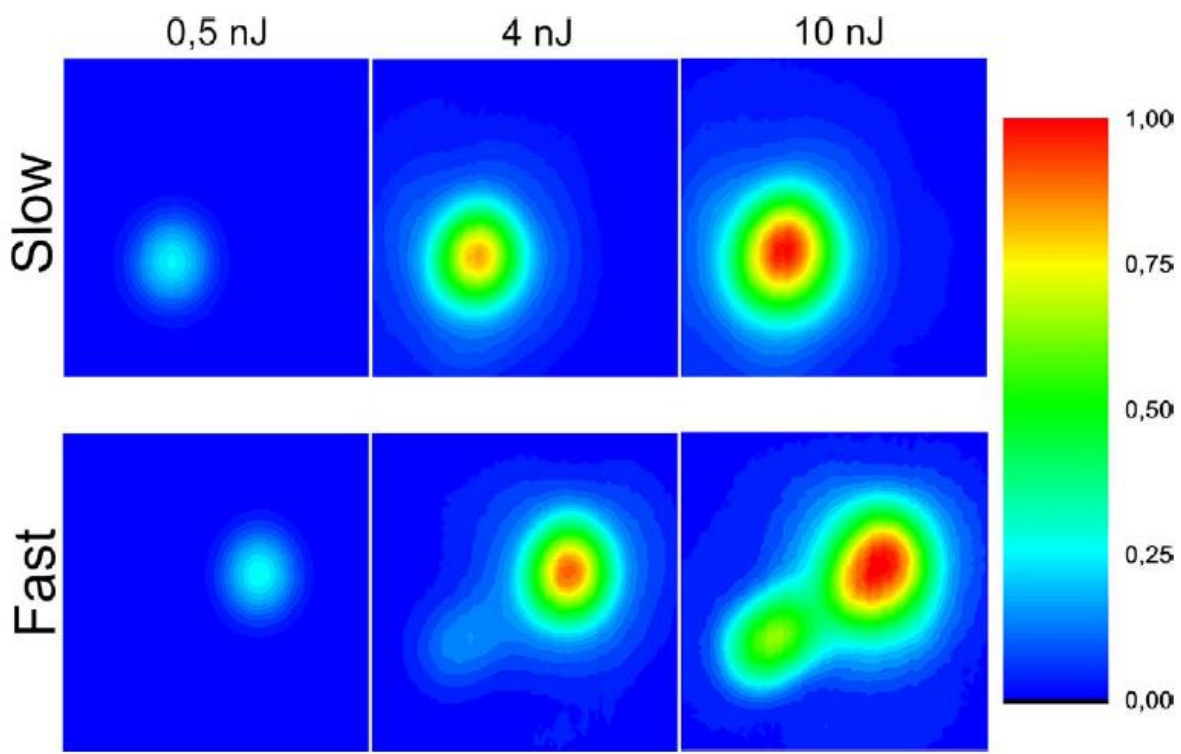

Fig. 3. Intensity distributions at the output facet of 10 mm long NL37D sample in the case of (top row) slow and (bottom row) fast core excitation by Y-polarized 1600 nm pulses at 0.5 nJ, 4 nJ and 10 nJ excitation pulse energies.

As it can be seen from the top row of Fig. 3, an increase of the excitation energy in-coupled into the slow core causes growth of the spatial intensity only in its own channel, without observable radiation exiting from the non-excited core. The absence of radiation at the output of non-excited core at all applied excitation energies and at both polarization directions indicates a negligible coupling between the cores under slow core excitation. In contrast, in the case of energy increase under fast core excitation (bottom row of Fig. 3), an increasing amount of radiation appears at the output of non-excited core at pulse energies equal to 4 nJ or higher. At this excitation energy the soliton number $N$ is estimated to have the value of $N = 12$ according to the formula

$$N^2 = \frac{\gamma P_0 T_0^2}{|GVD|}, \quad (4)$$

where $\gamma$ is the nonlinear parameter, $P_0$ is the peak power, $T_0$ is the temporal width of the incident pulse and $GVD$ is group velocity dispersion. The values of these parameters in the case of 1600 nm, Y polarization at 4 nJ pulse energy are as follows: $\gamma = 1.462$ 1/W.m, $P_0 = 11264$ W, $T_0 = 56.7$ fs, $GVD = -363$ fs$^2$/mm. In accordance with the previously introduced theory, the high order soliton underwent significant compression during the initial phase of its propagation, which ends after 1-2 mm depending on the input pulse energy. During this phase the nonlinear balancing can work only for a very short distance, because the peak intensity changes rapidly and therefore only negligible energy is transferable to the non-excited core. After the soliton fission process, the situation changes significantly. The separated fundamental solitons already hold their peak intensities and when they are stabilized at a proper level, the way is open for the energy transfer. However, even in this case, just partial transfer process is expectable because only a minor number (typically 1-3) of the fundamental solitons from the large number of the whole soliton ensemble (determined by the soliton number $N$) fulfill the asymmetry balancing condition

$$n_{eff}^{slow\_core} = n_{eff}^{fast\_core} + n_{NL} I_{peak}, \quad (5)$$

where $n_{eff}^{slow\_core}$ and $n_{eff}^{fast\_core}$ represent effective refractive indices of the slow and fast core respectively and $I_{peak}$ is the peak intensity of in-coupled pulse.

Similar asymmetric behavior, as presented in Fig.3, in terms of excited core choice was observable in the majority of the measurement series regardless of polarization direction and excitation wavelength in the case of both fiber samples. A straightforward explanation for the observed behavior is disturbed coupling between the fiber cores due to nonlinearly-induced growth of asymmetry in the case of the slow core excitation. In contrast, there is a possibility of eliminating the asymmetry under fast core excitation even at moderate excitation pulse energies thanks to the high Kerr nonlinearity of the chosen fiber material. In the case of NL37C fiber, the camera registrations exhibit similar behavior, however with lower measure of dual-core extinction ratio change. This finding is explainable by higher initial asymmetry of the fiber (according to the values presented in Table 2) and larger size of the cores resulting in a smaller nonlinearity and consequently weaker nonlinear asymmetry balancing effect.

The output spectra recorded from individual fiber cores under the experimental conditions corresponding to IR camera measurements presented in Fig. 3, i.e. excitation of 10 mm long NL37D sample with Y-polarized 1600 nm pulses, are depicted in Fig. 4.

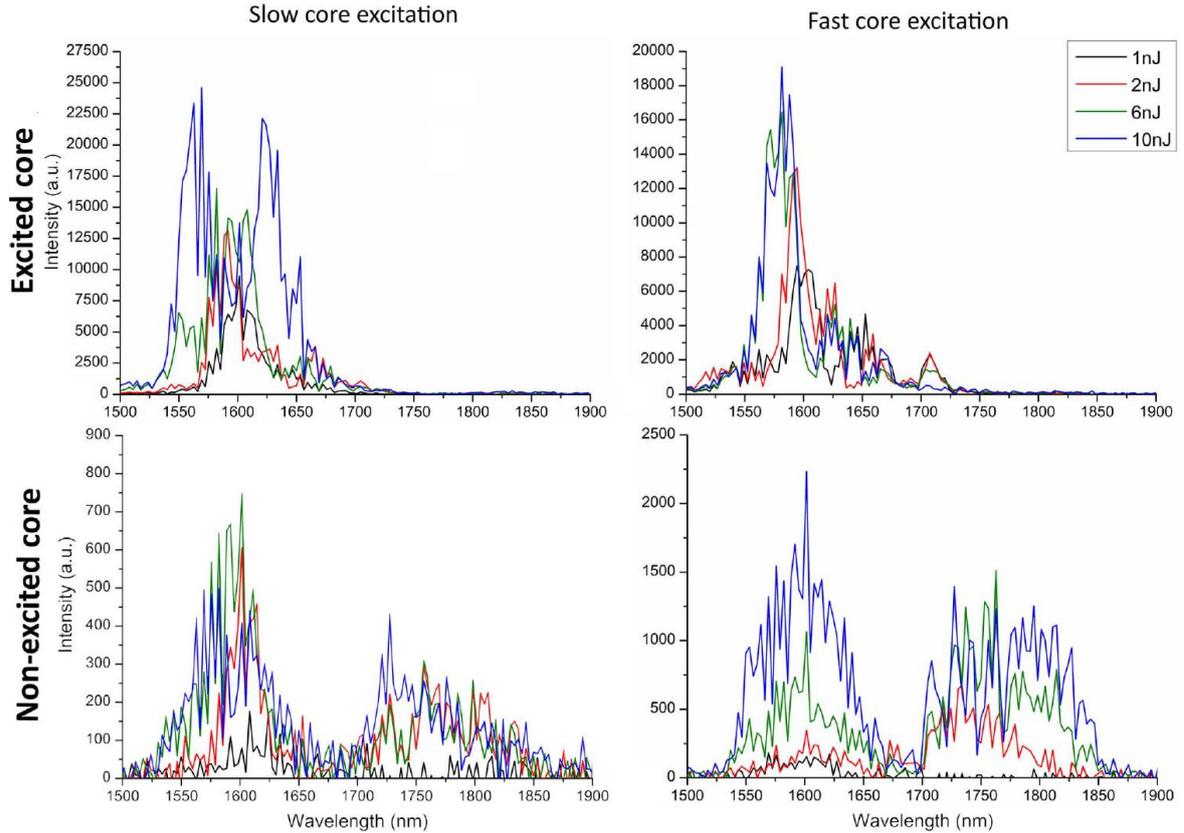

Fig. 4. Excitation energy dependent output spectra registered selectively from (top row) excited and (bottom row) non-excited core in the cases of excitation of (left column) slow vs. (right column) fast core. The measurements were performed under excitation of 10 mm long NL37D sample with Y-polarized 1600 nm pulses.

According to the graphs presented in Fig. 4 (top), the excited core output expresses lower spectral intensities and narrower bandwidths under fast core excitation in comparison to the slow one. It is typical asymmetric coupler behavior causing transfer of larger energy amount into the non-excited core under fast core excitation [18] as can be also seen from the graphs in Fig. 4 (bottom). This observation is in correspondence with the theory predicting higher probability of energy transfer from the fast core to the slow one in comparison to the opposite process [4]. However, a non-trivial feature is the evolution of the red-shifted spectral band around 1775 nm in the non-excited core separated from the fundamental spectral band. In the case of fast core excitation, this feature is absent in the excited core, but it has comparable intensity as the fundamental band in the non-excited one. It is explainable by the results of numerical simulations obtained under identical excitation conditions presented in Fig. 5. Compared to experimentally measured spectra, the simulated spectra are more structured. This is a common effect for this type of study [19] caused by the intensity fluctuations of the laser source, which leads to smoothing of the experimentally recorded spectra due to the inevitable time-integration. On the other hand, the simulated spectra span over wider spectral range than the experimental ones, indicating overestimation of the nonlinear refractive index during the simulation work.

Nevertheless, the simulation results provide the same qualitative picture about the ultrafast solitonic propagation in asymmetric coupler like the experimental ones. In the case of fast core excitation, higher energy transfer ratio to the non-excited core is predicted together with the energy-induced red-shift. The second confirmed feature is the pronounced long wavelength part of the spectrum in the non-excited core in comparison to the excited one. According to the all acquired simulation data, the origin of the red shifted band is the propagation phase right after the soliton fission process when some of the fundamental solitons propagate together separated from each other in the time domain. There is some leak of their spectral intensity from the excited channel to the non-excited one even under slow core excitation. Compared to the excited core, roughly two order of magnitude weaker intensities can be observed at the output from the non-excited one, both in experimental and simulation results. The relative weight of the red part of the spectra in the non-excited core is always higher because the coupling efficiency increases with increasing wavelength (Fig. 1b). However, under fast core excitation, the nonlinear interaction enhances the relative weight of the red part of the spectra in the non-excited core even more. In this case some redshifted fundamental solitons balance the effective refraction indices of the two channels and therefore transfer their energy to the non-excited one.

The transfer process is supported by the fact, that the fundamental solitons maintain their duration in the later propagation phase, both in the excited and non-excited core. However, after the transfer process, soliton remains trapped in the non-excited core due to the prohibited backward transfer. The other fundamental components generated simultaneously remain dominantly in the excited core because they have different peak intensities. Thus they do not fulfill precisely the asymmetry balancing condition (5), so only partial switching takes place according to both experimental and simulation results. It is worth mentioning that the key importance transfer processes evolve directly after the soliton compression point according to the simulation results, so we expect similar spatial-spectral field redistribution already at the level of 3 mm fiber length. The fast transfer process is caused by the short coupling length, which is at the level of 0,47 mm at Y-polarized 1600 nm excitation. The fast transfer concept is supported also by our experimental results obtained studying 6 mm length samples, expressing almost the same field redistribution between the two cores as in the case of 10 mm sample. Below 6 mm is really hard to manipulate with the fiber samples, therefore the fiber were not shortened more in the frame of the experimental study. At 1600 nm and Y-polarization the nonlinear length of our fiber is only 61 $\mu$m, more than two orders of magnitude less in comparison to the dispersion length (8.9 mm). Therefore the nonlinear asymmetry balancing is the dominant process under the studied conditions determined by the fiber characteristics and excitation wavelength. Of course, the characteristic length of the nonlinear balancing transfer (2-3 mm) is affected also by the soliton compression phenomena generating proper peak powers after significantly longer distances than the nonlinear length.

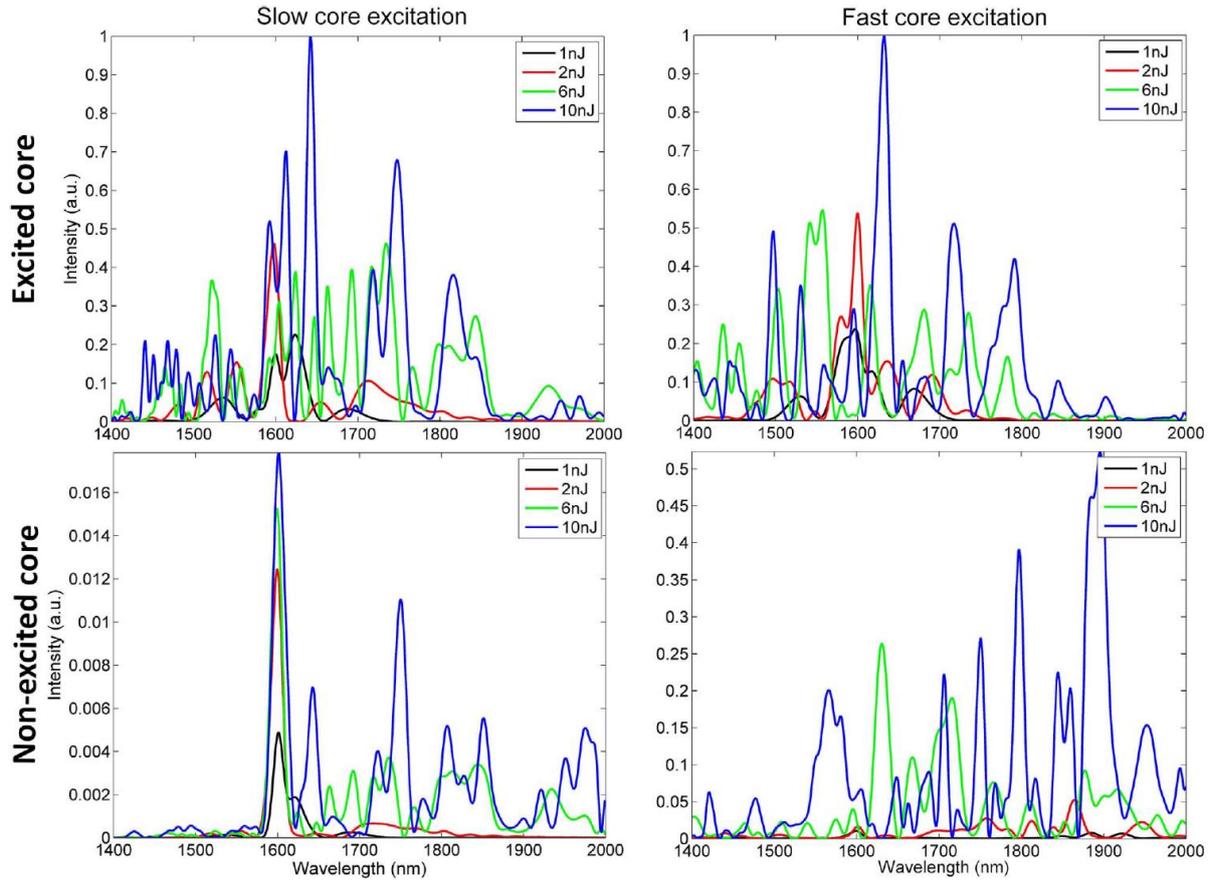

Fig. 5. Simulated excitation energy dependent output spectra of the (top row) excited and (bottom row) non-excited core in the cases of excitation of (left column) slow and (right column) fast core. The simulations were performed for 10 mm long NL37D sample at 1600 nm excitation wavelength and Y-polarization.

For better illustration of the studied phenomenon, input pulse energy dependence of energy transfer to the non-excited core was constructed (Fig. 6) by processing the spectral curves. The dependence represents the percentage ratio of integral energy transferred to the non-excited core ($E_{non-exc}$) with respect to overall integral energy in the both cores ($E_{exc} + E_{non-exc}$). This method of data processing revealed the monotonic increase (decrease) of energy transfer ratio from the excited to the non-excited core in the case of fast (slow) core excitation above 2 nJ. Both monotonic dependences confirm dominant role of nonlinearly-induced propagation constant change, which alters the energy transfer process and results in opposite behavior under excitation of the fast versus slow core. Due to the complex nature of high order solitonic propagation accompanied by significant soliton fission, the transfer takes place only for some fundamental soliton components when their peak intensities are optimal for the asymmetry balancing. Therefore, the proportional energy ratio remained below 50 % during all experimental series in the frame of this study. However, corresponding simulations under fast core excitation predict slightly higher transfer ratios accompanied by slight saturation character caused by higher influence of longer wavelength features (right column of Fig. 5) having higher values of $\kappa_e$ in comparison to the experimental spectra.

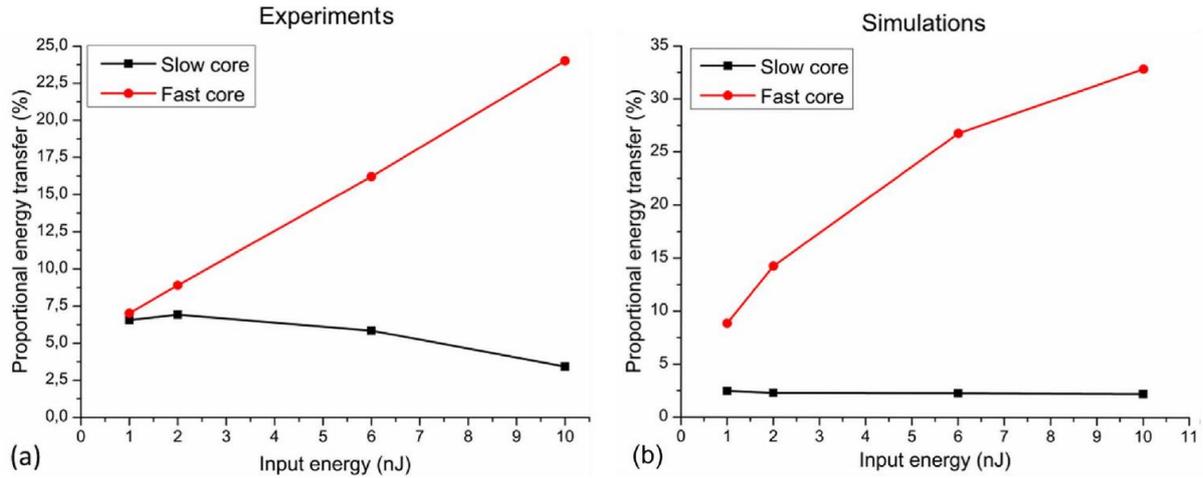

Fig. 6. (a) Experimental and (b) simulated proportional energy transfer from the excited to the non-excited core as a function of the input energy under slow and fast core excitation. 10 mm long NL37D sample was excited with 1600 nm, Y-polarized pulses.

*3.2 Impact of varying the excitation wavelength*

In the following we examine the influence of excitation wavelength on the nonlinearly-induced transfer effectiveness under fast core excitation. In Fig. 7 we present results obtained at two extreme excitation wavelengths of 1500 nm and 1800 nm, acquired in the case of 10 mm long NL37D sample under X-polarized excitation. According to the propagation constant mismatch listed in Table 2 the X-polarized radiation experiences weaker asymmetry in comparison to the Y-polarized one in the whole spectral range of our interest. Lower pulse energies are required in this case for the nonlinear asymmetry balancing, thus excitation with X-polarized pulses offers better conditions for the study of excitation wavelength effect.

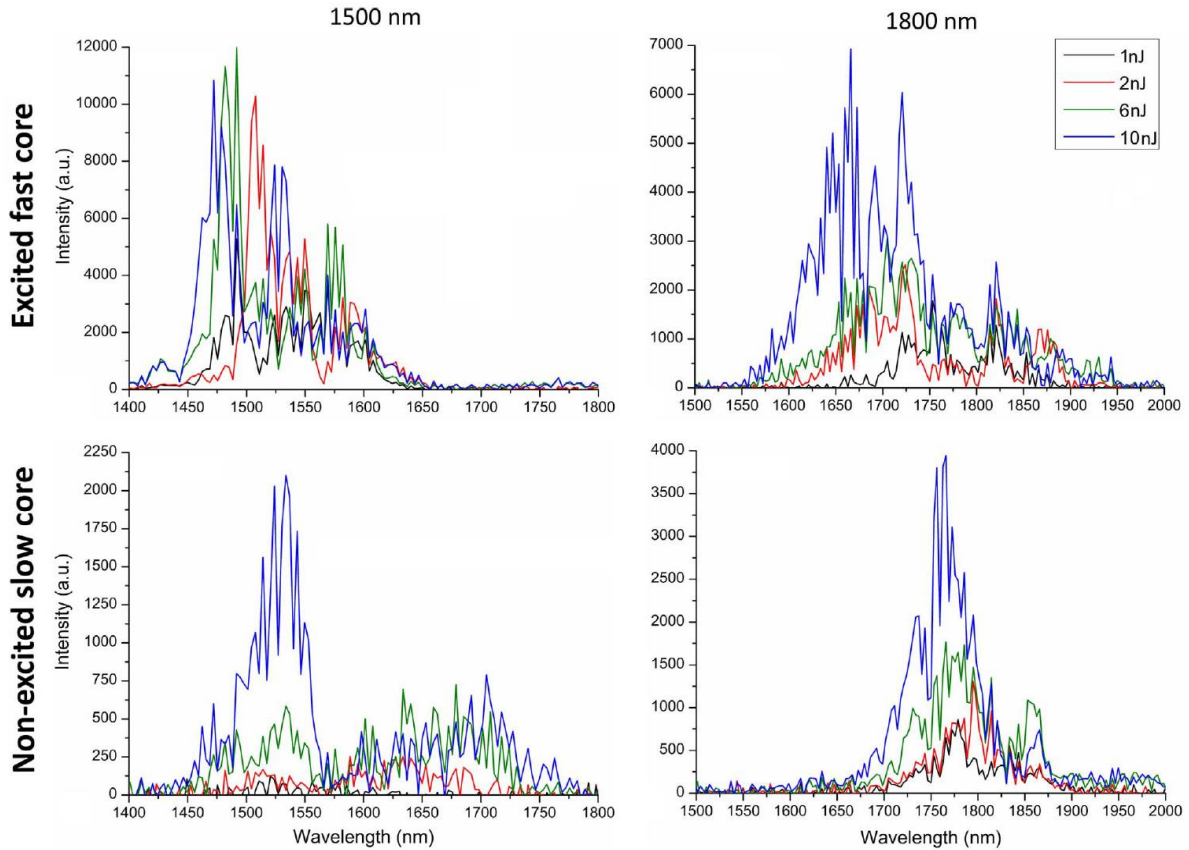

Fig. 7. Output spectra registered at different excitation pulse energies selectively from (top row) excited and (bottom row) non-excited core in the cases of (left column) 1500 nm and (right column) 1800 nm fast core excitation of 10 mm long NL37D sample with X-polarized pulses.

As it follows from the Fig. 7, comparing the spectral intensities between excited and non-excited core for both wavelengths, the investigated fiber supports significantly better coupling between the fiber cores at 1800 nm. This is true already for the linear interaction regime, because longer wavelengths support higher field overlap between the cores decreasing the measure of the asymmetry. As a result, lower peak intensities are sufficient for nonlinear asymmetry balancing. 1 nJ pulse energy already induces slight broadening in both cases, however from point of view of coupling the inter-core interaction is nearly the same as in the linear propagation region. The results show significantly higher energy transfer ratio in the case of the 1800 nm excitation with value of 25 % already at 1 nJ pulse energy (Fig. 8a), which is the best experimental result obtained at this energy level. Moreover, the initial value monotonically increases with input pulse energy in the case of both excitation wavelengths. Beside the more effective transfer, 1800 nm excitation wavelength supports also evolution of broader spectra in the excited core, mainly at the blue side, in comparison to the 1500 nm excitation.

In contrary, the non-excited core expresses broader spectra with two band character under 1500 nm excitation similarly to the previously analyzed 1600 nm excitation results. This observation is straightforward effect of the spectral dependence of $\kappa_e$ relaxing the peak-power requirement of the energy transfer with increasing wavelength (Fig. 1b). As we described above, the transfer process takes place dominantly after the soliton compression point and the key condition enabling the energy transfer is the suitable peak intensity of the fundamental solitons occurring at this propagation phase. According to the numerical simulations, in the case of the 1500 nm the red shifting and in the case of the 1800 nm the blue shifting processes dominate caused mainly by the opposite sign of the third order dispersion at these wavelengths (Fig. 8c). However, under 1800 nm excitation blue shifted fundamental solitons do not match the peak power condition of asymmetry balancing (5) and therefore they are missing in the non-excited core. On the other hand the components around the central wavelength are transferred with higher rates mainly due to the relaxed peak-power requirement at this wavelength. Under 1500 nm excitation the same phenomenon enhances the red part of the spectra in the non-excited core and causes broader overall width in comparison to the longer wavelength excitation.

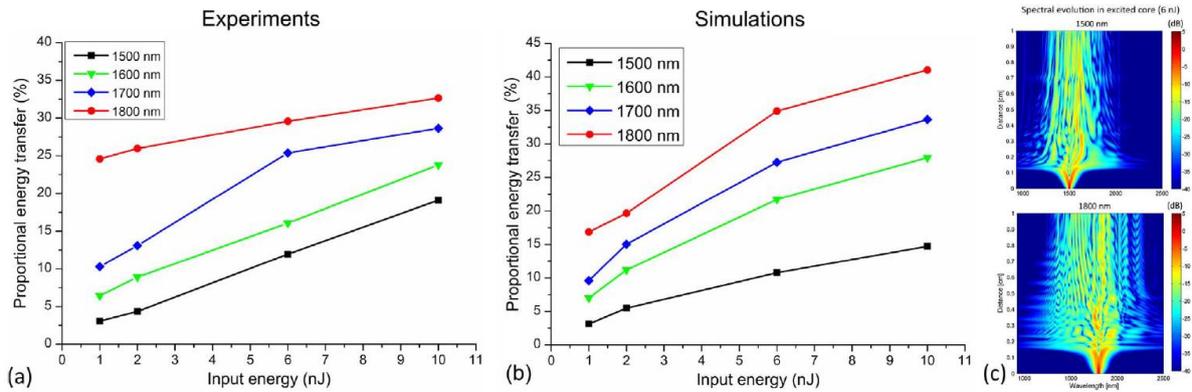

Fig. 8. (a) Experimental, (b) simulated input energy dependence of proportional energy transfer from excited to non-excited core in the case of different excitation wavelengths in range of 1500 - 1800 nm and (c) spectral evolution in dB in excited core under 1500 nm and 1800 nm excitation applying energy of 6 nJ (10 mm of NL37D sample, X-polarization).

Comparison of experimental proportional energy transfer curves for all studied excitation wavelengths in Fig. 8a reveals the consistence of the excitation wavelength effect. The transfer ratio monotonically increases with wavelength elongation already at 1 nJ pulse energy level as a result of increasing efficiency of linear coupling between the cores, even though there is already a minor nonlinear interaction. The initial coupling efficiency is enhanced further nonlinearly with increasing excitation pulse energy for all examined wavelengths with the best transfer ratios in the case of 1800 nm. The advantage of the increasing $\kappa_e$ with increasing wavelength (Fig. 1b) is obvious according to the whole dataset. The energy transfer ratio is enhanced by wavelength elongation in the case of the all applied pulse energy levels. The simulation results presented in Fig. 8b reflect the same character both in terms of energy and wavelength dependence with slightly higher transfer ratios which were noticed already in the introducing case of 1600 nm excitation. The monotonic increase of the transfer ratio implies the potential of its further enhancement by applying even higher pulse energies. However, further increase of input pulse energy during our experiments was not feasible due to the damage of the fibers observed above 10 nJ pulse energies.

### 3.3 High-contrast nonlinear switching

Motivated by the above presented findings, detailed switching analysis of the acquired spectra was performed for both NL37C and NL37D samples under fast core excitation at 1800 nm wavelength. The analysis consists of evaluation of extinction ratio (*ER*) between the spectral intensities recorded from the individual cores in decibels (dB) at all applied pulse energies (*E*) according to the formula

$$ER(E,\lambda) = 10\log\left(\frac{I_{excited}(E,\lambda)}{I_{non-excited}(E,\lambda)}\right). \quad (6)$$

Positive and negative values of the ER indicate dominance of the excited and non-excited core, respectively. The 3D color map in Fig. 9c is the joint representation of the wavelength and energy dependence of *ER* in conditions when the best nonlinear switching performance was found: 10 mm long NL37C sample, Y-polarized pulses at 1800 nm.

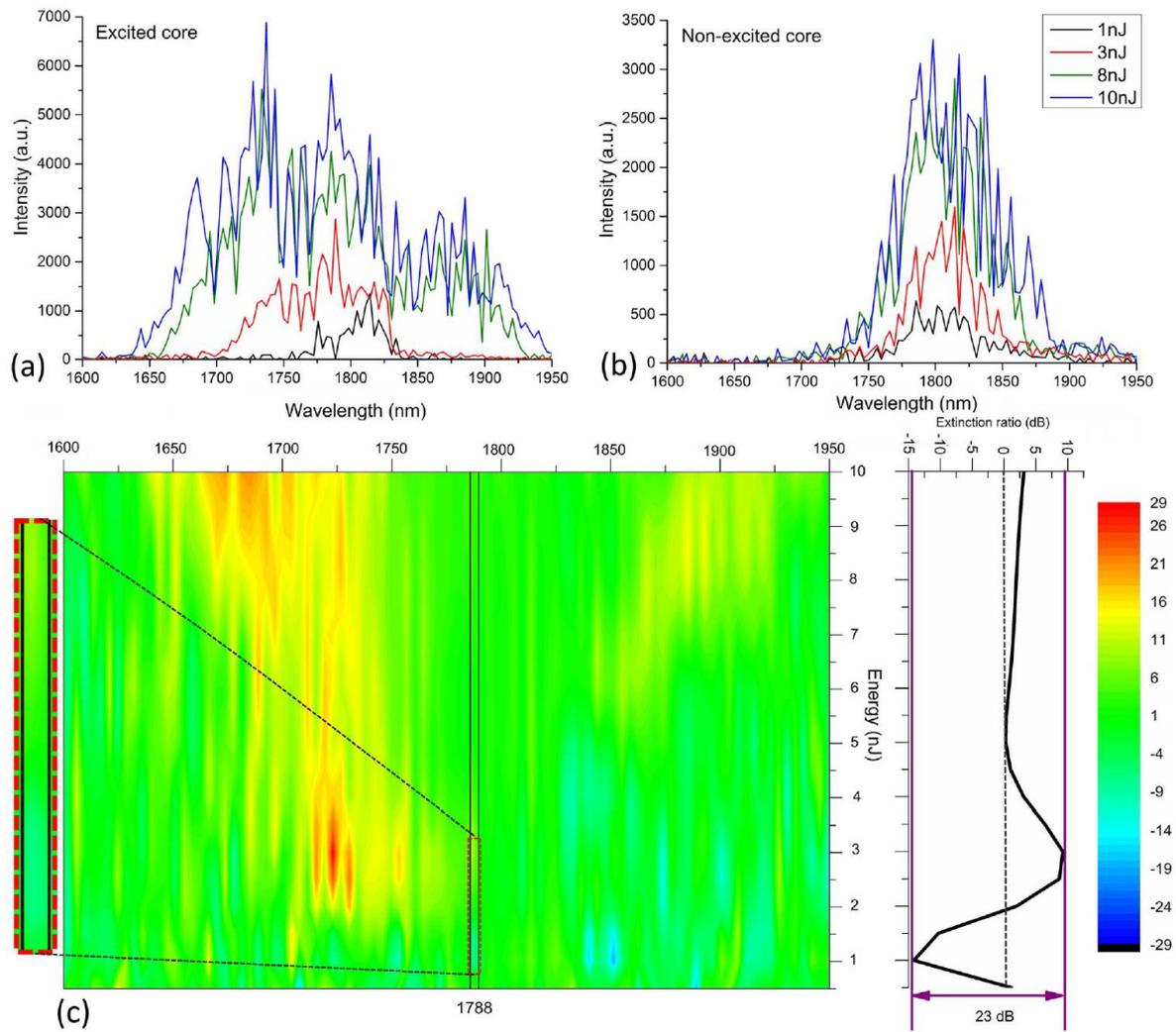

Fig. 9. Output spectra originating from (a) excited, (b) non-excited core at selected input energies and (c) 3D map of extinction ratio values depending on wavelength and input energy. The inset on the right side of (c) represents the extinction ratio dependence on input energy for selected wavelength 1788 nm. 10 mm long NL37C sample excited by Y-polarized pulses at 1800 nm.

As it can be seen from the results, strong spectral broadening dominating on the blue side of the spectra was observed in the excited core, similarly like in the case of the NL37D sample (right column of Fig. 7). In contrast, significantly narrower spectral feature centered around excitation wavelength was observed in the non-excited core. This again expresses enhancement of the transfer of fundamental solitonic components only around the central wavelength. As the resulting spectral shape is formed by complex soliton-based nonlinear processes, many narrow-band spectral features arise at higher input pulse energies. Therefore it is possible to find some narrow spectral bands, where the dominance is exchanged between the output cores, when two proper pulse energies are applied. This change of dominance represents nonlinear dual-core switching. However, in the majority of cases, the switching effect occurs at the periphery of the spectra where the spectral intensities are too low from practical point of view. It is clearly observable also in Fig. 9 containing many wavelengths at which the extinction ratio map changes its color from blue to yellow/orange in vertical direction, i.e. the signature of the *ER* switches from negative to positive. However, in few cases of the whole ensemble of the obtained results, nonlinear dual-core switching performance was identified around the central wavelength bearing significant spectral intensities.

The best recorded result, which is presented in Fig. 9 expresses narrow band switching in vicinity of wavelength 1788 nm with extinction ratio difference of 23 dB by changing the input pulse energy from 1 to 3 nJ. Taking into account the measured in-coupling efficiency of $C = 32\%$, the switching was achieved by in-coupled energy change from 0.32 to 0.96 nJ, thus in sub-nanojoule energy range. Positive aspect of this outcome is that extreme values of the *ER* are symmetrically displaced in relation to the zero level (shown also in Fig. 9c), so that applicable optical signal exits from the dominant channel at both switching energies. Moreover, 23 dB value represents large progress in comparison to our previously reported results exhibiting symmetric switching behavior at the level of 15 dB [19].

On the other hand, in the case of the presented results, switching takes place in narrower spectral area (in the vicinity of presented wavelength 1788 nm ± 3 nm) containing only a small part of launched excitation energy. Strongly modulated spectra and narrower features are caused by higher soliton number (N~10 at 1 nJ in-coupled pulse energy), determined mainly by high nonlinear refractive index of the utilized glass. Considering the enhancement of $n_{NL}$ by 20-times, we expected even smaller switching energies in the case of totally symmetric structure [20]. However, in the real asymmetric structure, energies approaching 1 nJ are required to fulfill the nonlinear asymmetry balancing condition (5) in order to obtain true switching performance. These energies correspond to higher soliton numbers than in our previous works [18, 19], which is drawback due to the stronger effect of the soliton fission process. Despite that, recent results represent important step towards solitonic high-contrast broadband switching based on dual-core optical fiber with high application potential. The high-contrast switch observed around the central wavelength was never demonstrated before in the case of solitonic regime of DC PCF excitation. Moreover, used numerical model brought comparable results , thus its capability for simulation of complex nonlinear dual-core propagation was confirmed also in the case of significant dual-core asymmetry.

The most important achievement of the presented study is that we successfully realized DC PCF based ultrafast switching at one order of magnitude lower pulse energies in comparison to the previous results. This was achieved mostly thanks to implementing highly nonlinear glass as fiber material. It is important to stress out that the geometrical parameters of the fiber were optimized supposing ideally symmetrical structure. However, the required level of fiber symmetry was not reached due to the challenging fabrication technology of this special type of highly nonlinear glass. Unfortunately the slight geometrical asymmetry was enhanced by high refraction index of the glass causing too high propagation constant difference between the cores. In spite of this drawback, the nonlinear asymmetry balancing principle was successfully confirmed by the systematic experimental study supported by the dedicated numerical simulations. Moreover, spectral intensity switching around central wavelength was demonstrated with dual-core extinction ratio change exceeding 20 dB level, what indicates the possibility to enhance our previous achievements. From application point of view, further goal is to work at this high-contrast switching level in broader spectral area, which can be achieved by decreasing the dual-core asymmetry and applying lower pulse energies.

## 4. Conclusions and outlook

Ultrafast all-optical switching based on nonlinear asymmetry balancing was demonstrated in self-designed DC PCF made of highly nonlinear lead-silicate glass. Effect of the input port change of the dual-core fiber was studied by simple monitoring of the output ports using the IR camera. The camera records revealed dissimilar character of the energy transfer between the cores in these two cases. Consistently, the IR camera and spectrometer recordings of the output field confirmed higher energy transfer rate to the non-excited core in case of fast core excitation. The key role of the fiber asymmetry in the spectral intensity redistribution between the two output ports was confirmed by monotonic dependences of the proportional energy transfers to the non-excited core on the excitation pulse energy at all excitation wavelengths in region of 1500 – 1800 nm. Analyzing the excitation energy dependences under fast core excitation, the longest applied wavelength of 1800 nm resulted in the highest energy transfer ratios. Obtained results were interpreted by the spectral dependence of the effective coupling coefficient between the slow and fast core.

Furthermore, the complex non-trivial nature of spectral distributions between the core outputs depending on the excitation wavelength was successfully addressed by numerical simulations based on the generalized CNLSE. Careful numerical analysis revealed the key mechanisms responsible for different measure of energy transfer ratios in various areas of the nonlinearly broadened spectra. Detailed switching analysis of both investigated samples under optimal 1800 nm excitation was performed. In the best case, 23 dB switching contrast close to the excitation wavelength was achieved at applicable spectral intensity levels using 10 mm long DC PCF sample and switching energies of 0.32 vs. 0.96 nJ. These values represent an order of magnitude reduction of the required switching energies at simultaneous improvement of the switching contrast and the required fiber length in comparison to the previous work. Obtained results constitute an important step towards applicable ultrafast DC PCF switch at the sub-nJ and sub-cm area in terms of required energies and fiber lengths, respectively. Moreover, they extend the validity of the employed numerical simulations model for the case of asymmetric couplers. By reduction of the fiber asymmetry one can reduce switching energies and unwanted high order soliton fission process in order to demonstrate broadband highly effective switching device. There is a realistic way to achieve this goal using the all-solid PCF technology by combining a highly nonlinear glass PBG08 with thermally matched low index glass [27, 28]. The development of such a new DC PCF sample with appropriate linear and nonlinear characteristics constitutes our future research plan, which however lies already beyond the scope of the presented study.

*Acknowledgements*
The authors acknowledge the support of Austrian Science Fund (FWF) [P 27577-N27], Foundation for Polish Science co-financed by the European Union under the European Regional Development Fund [TEAM TECH/2016-1/1], Ernst Mach Grant - Action Austria-Slovakia [ICM-2015-00653], Grant agency of the Ministry of Education of the Slovak Republic [VEGA 1/0929/17], International Laser Centre in Bratislava and Comenius University, Faculty of Mathematics, Physics and Informatics.

*Funding sources*
This work was supported by the Foundation for Polish Science co-financed by the European Union under the European Regional Development Fund [TEAM TECH/2016-1/1], Ernst Mach Grant - Action Austria-Slovakia [ICM-2015-00653], Austrian Science Fund (FWF) [P 27577-N27] and and National Science Centre, Poland [project No. 2016/23/P/ST7/02233 under POLONEZ programme which has received funding from the European Union's Horizon 2020 research and innovation programme under the Marie Skłodowska-Curie grant agreement No 665778].

**Footnotes list:** [1] List of abbreviations

**References**
[1]   P.S.J. Russell, Photonic-crystal fibers, J. Light. Technol. 24 (2006) 4729–4749. doi:10.1109/JLT.2006.885258.
[2]   K. Saitoh, Y. Sato, and M. Koshiba, "Coupling characteristics of dual-core photonic crystal fiber couplers," Opt. Express **11**, 3188-3195 (2003), doi: 10.1364/OE.11.003188
[2]   A. Betlej, S. Suntsov, K.G. Makris, L. Jankovic, D.N. Christodoulides, G.I. Stegeman, J. Fini, R.T. Bise, D.J. Digiovanni, All-optical switching and multifrequency generation in a dual-core photonic crystal fiber., Opt. Lett. 31 (2006) 1480–1482. doi:10.1364/OL.31.001480.

[3]     G.P. Agrawal, Applications of Nonlinear Fiber Optics, Academic Press, New York, 2001.
[4]     S. Trillo, S. Wabnitz, Nonlinear nonreciprocity in a coherent mismatched directional coupler, Appl. Phys. Lett. 49 (1986) 752–754. doi:10.1063/1.97536.
[5]     L. Curilla, P. Stajanca, I. Bugar, R. Buczynski, F. Uherek, Dual-core microstructure optical fiber as a potential polarization splitter, Proc. SPIE - Int. Soc. Opt. Eng. 9441 (2014) 1–11. doi:10.1117/12.2087316.
[6]     H. Jiang, E. Wang, J. Zhang, L. Hu, Q. Mao, Q. Li, K. Xie, Polarization splitter based on dual-core photonic crystal fiber, Opt. Express. 22 (2014) 30461–30466.
[7]     S. Trillo, S. Wabnitz, E.M. Wright, G.I. Stegeman, Soliton switching in fiber nonlinear directional couplers., Opt. Lett. 13 (1988) 672–674. doi:10.1364/OL.13.000672.
[8]     T. Uthayakumar, R. Vasantha Jayakantha Raja, K. Nithyanandan, K. Porsezian, Designing a class of asymmetric twin core photonic crystal fibers for switching and multi-frequency generation, Opt. Fiber Technol. 19 (2013) 556–564. doi:10.1016/j.yofte.2013.08.010.
[9]     A.K. Sarma, Soliton switching in a highly nonlinear dual-core holey fiber coupler, Jpn. J. Appl. Phys. 47 (2008) 5493–5495. doi:10.1143/JJAP.47.5493.
[10]    W.B. Fraga, J.W.M. Menezes, M.G. da Silva, C.S. Sobrinho, A.S.B. Sombra, All optical logic gates based on an asymmetric nonlinear directional coupler, Opt. Commun. 262 (2006) 32–37. doi:10.1016/j.optcom.2005.12.033.
[11]    A.G. Coelho, M.B.C. Costa, A.C. Ferreira, M.G. Da Silva, M.L. Lyra, A.S.B. Sombra, Realization of all-optical logic gates in a triangular triple-core photonic crystal fiber, J. Light. Technol. 31 (2013) 731–739. doi:10.1109/JLT.2012.2232641.
[12]    D. Lorenc, I. Bugar, M. Aranyosiova, R. Buczynski, D. Pysz, D. Velic, D. Chorvat, Linear and nonlinear properties of multicomponent glass photonic crystal fibers, Laser Phys. 18 (2008) 270–276. doi:10.1007/s11490-008-3013-7.
[13]    A. V. Husakou, J. Herrmann, Supercontinuum Generation of Higher-Order Solitons by Fission in Photonic Crystal Fibers, Phys. Rev. Lett. 87 (2001) 203901. doi:10.1103/PhysRevLett.87.203901.
[14]    J. Herrmann, U. Griebner, N. Zhavoronkov, A. Husakou, D. Nickel, J.C. Knight, W.J. Wadsworth, P.S. Russell, G. Korn, Experimental evidence for supercontinuum generation by fission of higher-order solitons in photonic fibers, Phys. Rev. Lett. 88 (2002) 1739011–1739014. doi:10.1103/PhysRevLett.88.173901.
[15]    F. Luan, A. Yulin, J.C. Knight, D. V. Skryabin, Polarization instability of solitons in photonic crystal fibers, Opt. Express. 14 (2006) 6550. doi:10.1364/OE.14.006550.
[16]    I. Bugar, I. V Fedotov, A.B. Fedotov, M. Koys, R. Buczynski, D. Pysz, J. Chlpik, F. Uherek, A.M. Zheltikov, Polarization-Controlled Dispersive Wave Redirection in Dual-Core Photonic Crystal Fiber, Laser Phys. 18 (2008) 1420–1428. doi:10.1134/S1054660X08120086.
[17]    M. Koys, I. Bugar, I. Hrebikova, V. Mesaros, R. Buczynski, F. Uherek, Spectral switching control of ultrafast pulses in dual core photonic crystal fibre, J. Eur. Opt. Soc. 8 (2013) 13041. doi:10.2971/jeos.2013.13041.
[18]    P. Stajanca, D. Pysz, M. Michalka, G. Andriukaitis, T. Balciunas, G. Fan, A. Baltuska, I. Bugar, Soliton-based ultrafast multi-wavelength nonlinear switching in dual-core photonic crystal fibre, Laser Phys. 24 (2014) 1–10. doi:10.1088/1054-660X/24/6/065103.
[19]    P. Stajanca, D. Pysz, G. Andriukaitis, T. Balciunas, G. Fan, A. Baltuska, I. Bugar, Ultrafast multi-wavelength switch based on spectrally-shifted solitons dynamics in a dualcore photonic crystal fiber, Opt. Express. 22 (2014) 31092–31101. doi:10.1364/OE.22.031092.
[20]    P. Stajanca, I. Bugar, Nonlinear ultrafast switching based on soliton self-trapping in dual-core photonic crystal fibre, Laser Phys. Lett. 13 (2016) 1–8.
[21]    G. Sobon, M. Klimczak, J. Sotor, K. Krzempek, D. Pysz, R. Stepien, T. Martynkien, K.M. Abramski, R. Buczynski, Infrared supercontinuum generation in soft-glass photonic crystal fibers pumped at 1560 nm, Opt. Mater. Express. 4 (2014) 7–15. doi:10.1364/OME.4.000007.
[22]    P.L. Chu, B.A. Malomed, G.D. Peng, Soliton switching and propagation in nonlinear fiber couplers: analytical results, J. Opt. Soc. Am. B. 10 (1993) 1379–1385.
[23]    T. Kanna, M. Lakshmanan, Exact soliton solutions of coupled nonlinear Schrödinger equations: Shape-changing collisions, logic gates, and partially coherent solitons, Phys. Rev. E. 67 (2003) 46617. doi:10.1103/PhysRevE.67.046617.

[24] B.F. Feng, General N-soliton solution to a vector nonlinear Schrödinger equation, J. Phys. A Math. Theor. 47 (2014) 355203.

[25] M. Koys, I. Bugar, V. Mesaros, F. Uherek, R. Buczynski, Supercontinuum generation in dual core photonic crystal fiber, in: Proc. SPIE - Int. Soc. Opt. Eng., 2010: p. 11. doi:10.1117/12.882890.

[26] I. Cristiani, R. Tediosi, L. Tartara, V. Degiorgio, Dispersive wave generation by solitons in microstructured optical fibers, Opt. Express. 12 (2004) 124–135. doi:10.1364/OPEX.12.000124.

[27] X. Feng, T. Monro, P. Petropoulos, V. Finazzi, D. Hewak, Solid microstructured optical fiber., Opt. Express. 11 (2003) 2225–2230. doi:10.1364/OE.11.002225.

[28] M. Klimczak, B. Siwicki, P. Skibiński, D. Pysz, R. Stępień, A. Heidt, C. Radzewicz, R. Buczyński, Coherent supercontinuum generation up to 2.3 μm in all-solid soft-glass photonic crystal fibers with flat all-normal dispersion, Opt. Express. 22 (2014) 18824–18832. doi:10.1364/OE.22.018824.